\documentclass[a4paper,pdftex]{jpconf}
\usepackage{bm}
\usepackage{adjustbox}
\usepackage{graphicx}
\usepackage{hyperref}
\hypersetup{colorlinks,%
            citecolor=black,%
            filecolor=black,%
            linkcolor=black,%
            urlcolor=black,%
            pdftex}

\newcommand{\ud}     {\mathrm{d}}

\newcommand{\average}[1]{\left\langle{#1}\right\rangle}
\newcommand{\eq}[1]{Eq.(\ref{#1})}

 \newcommand{\sigmatot}{\sigma_\mathrm{tot}}
 \newcommand{\as}{\alpha_{\scriptscriptstyle S}}

\begin{document}

\title{Modelling the nuclear parton distributions}

\author{S A Kulagin}

\address{Institute for Nuclear Research of the Russian Academy of Sciences, Moscow 117312, Russia}

\ead{sergey.kulagin@gmail.com}

\begin{abstract}
We review a semi-microscopic model of nuclear parton distributions,
which takes into account a number of nuclear effects including
Fermi motion and nuclear binding, nuclear meson-exchange currents and off-shell corrections
to bound nucleon distributions as well as nuclear shadowing effect.
We also discuss applications of the model to the lepton-nuclear deep-inelastic scattering,
Drell-Yan process and neutrino total cross sections.
\end{abstract}

\section{Introduction}

The leading contribution to cross sections of various hard processes
is determined by the parton distributions (PDFs).
Thus, PDFs are universal process-independent characteristics
of the target at high invariant momentum transfer $Q$.
The nuclear parton distributions (NPDFs) are the subject of significant nuclear effects
which span a wide region of Bjorken $x$ as demonstrated by the deep inelastic scattering (DIS)
experiments
and the measurements of the Drell-Yan process (DY)
(for a review see \cite{Arneodo:1992wf,Norton:2003cb}).
A relative rate of nuclear effects is  more than
one order of magnitude larger than the ratio of the nuclear
binding energy to the nucleon mass thus indicating
that the nuclear environment plays an important role
even at high energies and momenta.

A number of phenomenological approaches to NPDFs are available in literature
\cite{Eskola:2009uj,Hirai:2007sx,deFlorian:2011fp,Kovarik:2015cma}.
Although such studies are useful in constraining nuclear effects
for different partons, they provide little information about the underlying physics
responsible for the nuclear corrections.
In this contribution we follow a different approach and present the results of a study of NPDFs
using a model of Ref.\cite{KP04} which takes into account a few different
mechanisms of nuclear corrections.
The model, which is briefly reviewed in Sec.\ref{sec:model},
explains the observed $x$, $Q^2$ and $A$ dependencies of
the nuclear structure functions in the deep inelastic scattering (DIS)
for a wide range of nuclear targets from $^2$H to ${}^{207}$Pb~\cite{KP04,KP07,KP10}.
In what follows we focus on the region of high $Q^2$ and review predictions \cite{KP14}
of nuclear effects for the valence and sea quark distributions (Sec.\ref{sec:model}),
discuss applications to nuclear DIS,
nuclear DY process (Sec.\ref{sec:ndy})
and the total neutrino-nuclear cross sections (Sec.\ref{sec:nuxsec}).

\section{Nuclear PDFs and DIS}
\label{sec:model}

We will use the notation $q_{a/T}(x,Q^2)$ for the distribution of quarks of the flavor
$a$ in a target $T$. The parton distributions in a nucleus
receive a number of contributions and can be summarized as \cite{KP04,KP14}
\begin{equation}
\label{npdf}
q_{a/A} = \average{q_{a/p}} + \average{q_{a/n}}
          + \delta q_a^\mathrm{coh} + \delta q_a^\mathrm{MEC} .
\end{equation}
For brevity, we suppress explicit dependencies on $x$ and $Q^2$ in \eq{npdf}.
The first two terms on the right side are the contribution from the partons from
bound protons and neutrons. The brackets stand for the averaging with the
energy-momentum distribution of bound nucleons in a nucleus,
as discussed in detail in Ref.\cite{KP04,KP14}.
Note that the evaluation of these terms requires the proton and the neutron PDFs
in off-mass shell region \cite{Kulagin:1994fz}.
The off-shell correction together with the nucleon momentum distribution (Fermi motion)
and the nuclear binding effect~\cite{FMB}
plays an important role in the valence quark region \cite{KP04,KP10}.

The correction $\delta q^\mathrm{coh}$
arises due to propagation effects of intermediate hadronic states
of a virtual boson in nuclear environment.
This term involves contributions from multiple scattering series
and typically lead to a negative correction  relevant at low $x$ (nuclear shadowing effect,
for a review see, e.g., Ref.\cite{Piller:1999wx}).

The last term in \eq{npdf} is a contribution from the meson degrees of freedom in nuclei.
We recall that the meson fields mediate the nucleon-nucleon interaction in nuclei
and also generate an additional quark-gluon content in nuclei.
This term is relevant for intermediate region of $x$ and results in a some enhancement
of the nuclear sea quark distribution.
Also its contribution is important to balance the overall nuclear light-cone momentum.
More details on the treatment of each term in \eq{npdf} can be found in Refs.\cite{KP04,KP14}.

In analysis of Ref.\cite{KP14} we choose the PDFs with definite $C$-parity $q^\pm = q \pm \bar q$ as a basis.
Note that the $C$-even combination $xq^+$ contributes to the structure function $F_2$,
for both the charged-lepton and neutrino deep-inelastic scattering, while the $C$-odd
$xq^-$ is the valence quark distribution which contributes to the neutrino structure function $xF_3$.
For  light quarks we also consider the isoscalar $q_0=u+d$ and the isovector $q_1=u-d$ combinations.
Below we summarize our results on the nuclear valence and sea quark PDFs.

\begin{figure}[htb]
\centering
\includegraphics[width=0.5\textwidth]{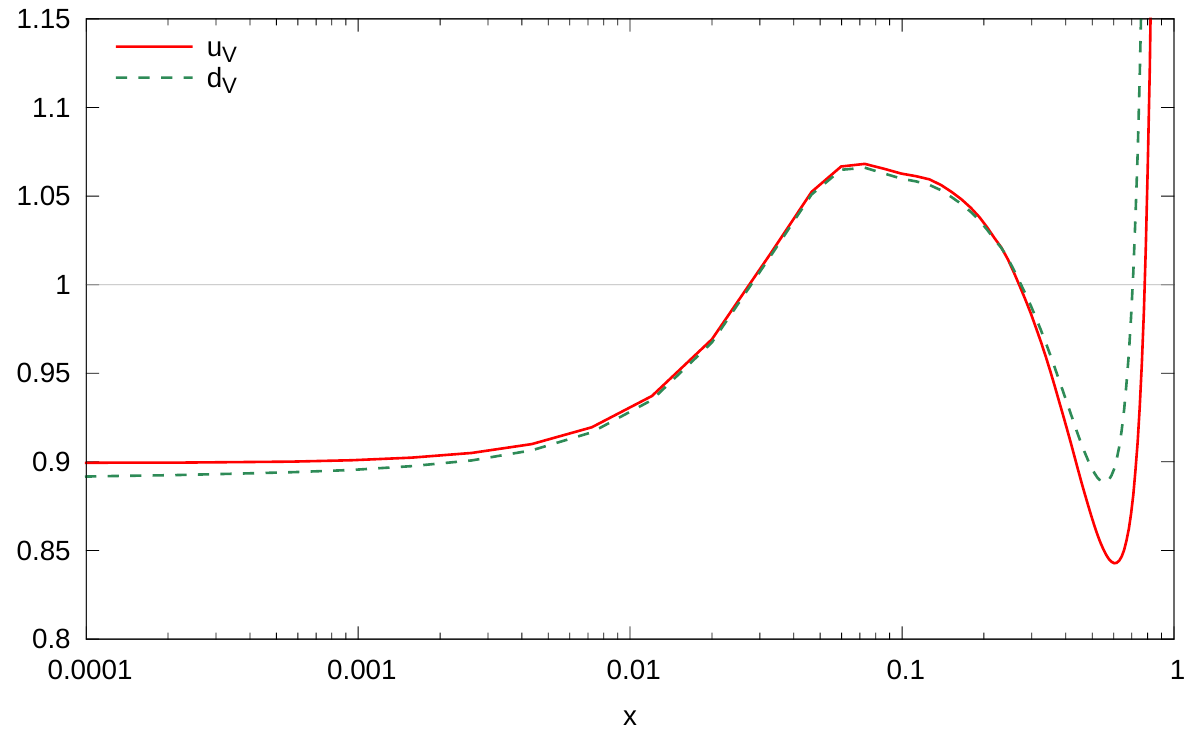}%
\includegraphics[width=0.5\textwidth]{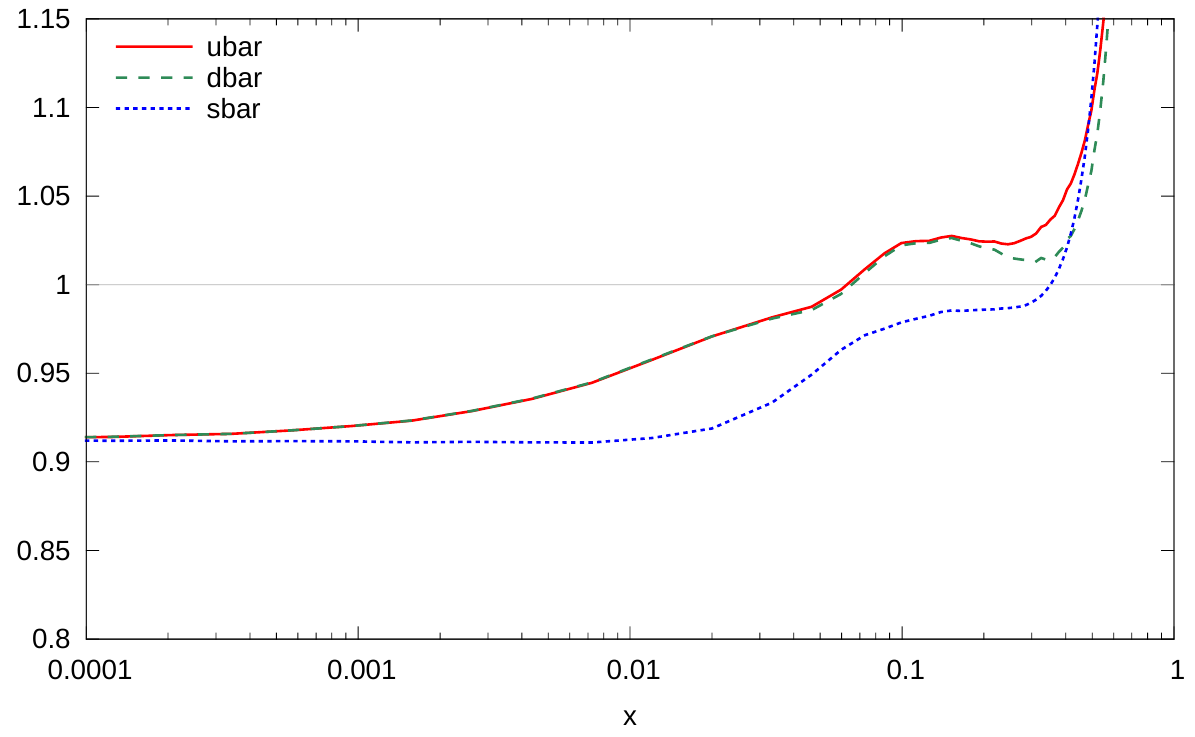}
\caption{%
The nuclear ratios for the valence quarks (left panel)
and the $\bar u$, $\bar d$, and $\bar s$ antiquarks (right panel)
in ${}^{208}$Pb calculated at $Q^2=25\, \mathrm{GeV}^2$
(see text for more detail).
\label{fig:npdfs}}
\end{figure}

In order to illustrate the nuclear effects, in  Fig.\ref{fig:npdfs} we show
the ratios $R_a^A=q_{a/A}/(Z q_{a/p}+N q_{a/n})$ calculated 
for the valence quarks (left panel) and the antiquarks (right panel) in the lead nucleus.
Here $Z$ and $N$ are the proton and the neutron number in the discussed nucleus with the total nucleon number $A=Z+N$,
and $q_{a/p}$ and $q_{a/n}$ are the free proton and the neutron PDF, respectively.
As can be seen from Fig.\ref{fig:npdfs} the resulting nuclear effect is different for the nuclear valence
and sea distribution and also depends on the PDF flavor.
We briefly comment on characteristic features of the nuclear ratios for different PDFs,
for a more detail discussion see Ref.\cite{KP14}.

In the region $x\ll 0.1$, all NPDFs are suppressed
due to the nuclear \emph{shadowing} effect (a negative $\delta q^{\rm coh}$ term).
Note that the rate of this effect is not universal and differs for
the valence and sea quark distribution \cite{KP04,KP14}.
We also remark that the result of averaging of the nucleon PDF
with the nuclear spectral function in the first two terms in the right side of \eq{npdf}
depends on the details of the $x$ dependence of considered PDF.
For this reason, this correction is positive for the valence quarks
thus partially reducing the shadowing effect.
For the sea quarks this correction is negative in this region.
However, the MEC correction is positive and we also have a partial cancellation of different effects.


In the intermediate region of $x\sim 0.1$
(which is usually referred to as \emph{antishadowing} region)
we observe interplay between different nuclear corrections.
For the valence quark PDF ($C$-odd $q^-$ distribution) we find an enhancement
which is due to constructive interference in the multiple scattering effect between
the $C$-even and $C$-odd amplitudes in the $\delta q^\mathrm{coh}$ term.
The antishadowing effect on  the antiquark PDFs is somewhat weaker and shifted towards larger $x$
because of a partial cancellation between different contributions \cite{KP14}.

At large $x>0.2$ the nuclear PDFs are dominated by incoherent scattering
from bound nucleons, i.e. by the first term on the right in \eq{npdf}.
For the valence quarks, the averaging of the proton and neutron PDFs with the nuclear spectral function
together with the off-shell correction produces a pronounced `EMC-effect' shape at large $x$ \cite{KP04}.
This effect was noticed long ago in \cite{FMB,Kulagin:1989mu} and further discussed in Ref.\cite{KP04}.
The relative rate of this correction strongly depends on the particular $x$ dependence of the given PDF
and for this reason the ratios $R_\mathrm{val}$ and $R_\mathrm{sea}$ are quite different at large $x$.

It should be remarked that different nuclear effects are related through a number of sum rules
which reflect different global symmetries.
In particular, the conservation of the valence quark number links together the off-shell and
the shadowing corrections to the valence quark distributions.
In Ref.\cite{KP04,KP14} the normalization conditions for the isoscalar and the isovector
valence quark distributions were used as
equations to determine the unknown amplitudes controlling the coherent nuclear correction.
We then solved these equations in terms of the off-shell correction to the corresponding
distributions.
The energy-momentum conservation causes the light-cone momentum sum rule at two different levels.
At the hadronic level, the nuclear light-cone momentum is shared between nucleons and mesons.
This requirement allows to constrain the MEC correction to NPDFs.
At the partonic level, the light-cone momentum is balanced between quarks, antiquarks and gluons.
The study of different contributions to the light-cone momentum sum rule can provide insights
on modification of gluon distribution in nuclei.

\begin{figure}[!]
  \begin{adjustbox}{addcode={\begin{minipage}{\width}}{\caption{%
    The ratio of a nucleus to the deuteron structure function $F_2^A/F_2^D$ for different nuclei.
    Data are from experiments at CERN (EMC \cite{EMC}, NMC \cite{NMC} and BCDMC \cite{BCDMS}),
    SLAC (E139 \cite{E139}), and JLab \cite{jlab}.
    The model predictions are given for a fixed $Q^2=5$\,GeV$^2$. 
    \label{fig:emc-effect}}\end{minipage}},rotate=90,center}
    \includegraphics[width=0.95\textheight]{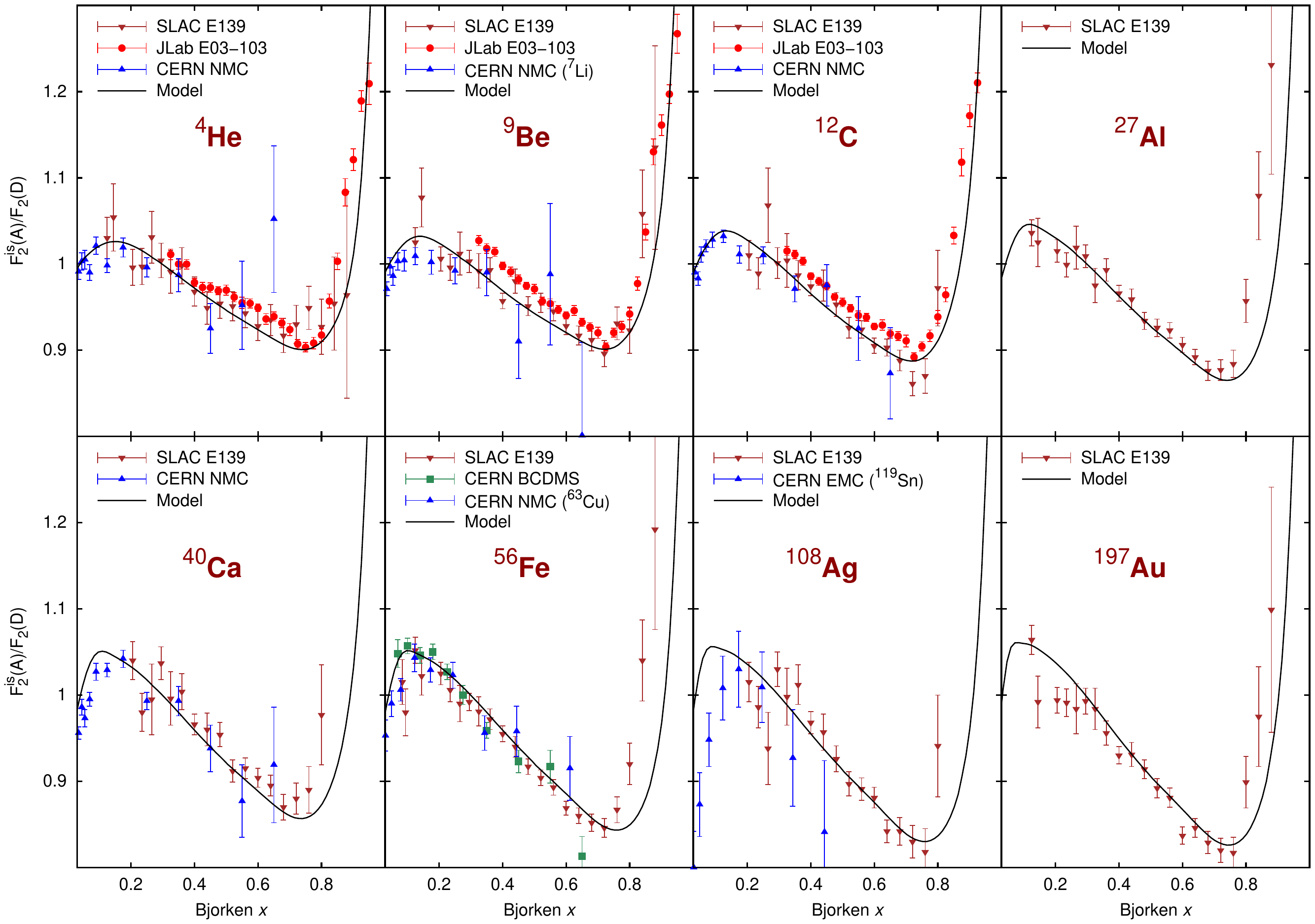}%
  \end{adjustbox}
\end{figure}

A detailed analysis of data on the ratio of the DIS structure functions for different nuclear targets
$R_2(A'/A)=F_2^{A'}/F_2^A$ (usually the ratio is taken with respect to the deuteron)
with targets ranging from $^4$He to $^{208}$Pb
shows a very good agreement of  model predictions with  observed $x$, $Q^2$ and
nuclear dependence of data \cite{KP04,KP10}.
The performance of discussed model is illustrated in Fig.\ref{fig:emc-effect}, in which we show
the model predictions in comparison with data for a number of nuclei.
However, it should be remarked that the model predictions in Fig.\ref{fig:emc-effect} are given at
a fixed $Q^2=5$\,GeV$^2$ while the data
from different measurements have essentially different average $Q^2$.
Furthermore, each $x$ bin of data has different average $Q^2$ value.
For a detailed comparison with data and discussion of consistency of different data sets see Ref.\cite{KP04,KP10}.

\section{NPDFs and Drell-Yan process}
\label{sec:ndy}

The reaction of muon pair production in hadron-hadron collisions (Drell-Yan process)
is an important source of information on the proton, pion and nuclear PDFs \cite{Peng:2014hta}.
In the context of NPDFs, the use of DY data in combination with DIS data allows a separation
of the nuclear valence and sea quark distributions.
In the DY reaction with the proton beam by tuning the kinematics of the muon pair
one can select a region in which the DY cross sections are driven by annihilation of valence
quarks in the beam and antiquarks in the target. Then the ratio of the DY cross sections
off different nuclear targets provide
a tool to measure the nuclear dependence of antiquark PDFs
\begin{equation}\label{eq:dy_r}
\frac{\sigma^{DY}_A}{\sigma^{DY}_B}
	\approx
	\frac{\bar u_A(x_T) + \bar d_A(x_T)}{\bar u_B(x_T) + \bar d_B(x_T)} ,
\end{equation}
where $x_T$ is the Bjorken variable for a nuclear target.
This ratio  was measured in the
experiments E772~\cite{E772} and E866~\cite{E866}
with the proton beam momentum 800 GeV/c at Fermilab in the region
$x_T < 0.15$ for a number of nuclear targets.
In contrast to DIS, the DY  data show no \emph{antishadowing} (i.e. enhancement of nuclear antiquark distributions)
at $x_T\sim 0.1$ that was a long standing puzzle since the nuclear
binding should result in an
excess of nuclear mesons, which is expected to produce a marked enhancement in the nuclear
anti-quark distributions \cite{Bickerstaff:1985ax}.
\begin{figure}[htb]
\centering
\vspace{-1em}\includegraphics[width=\textwidth]{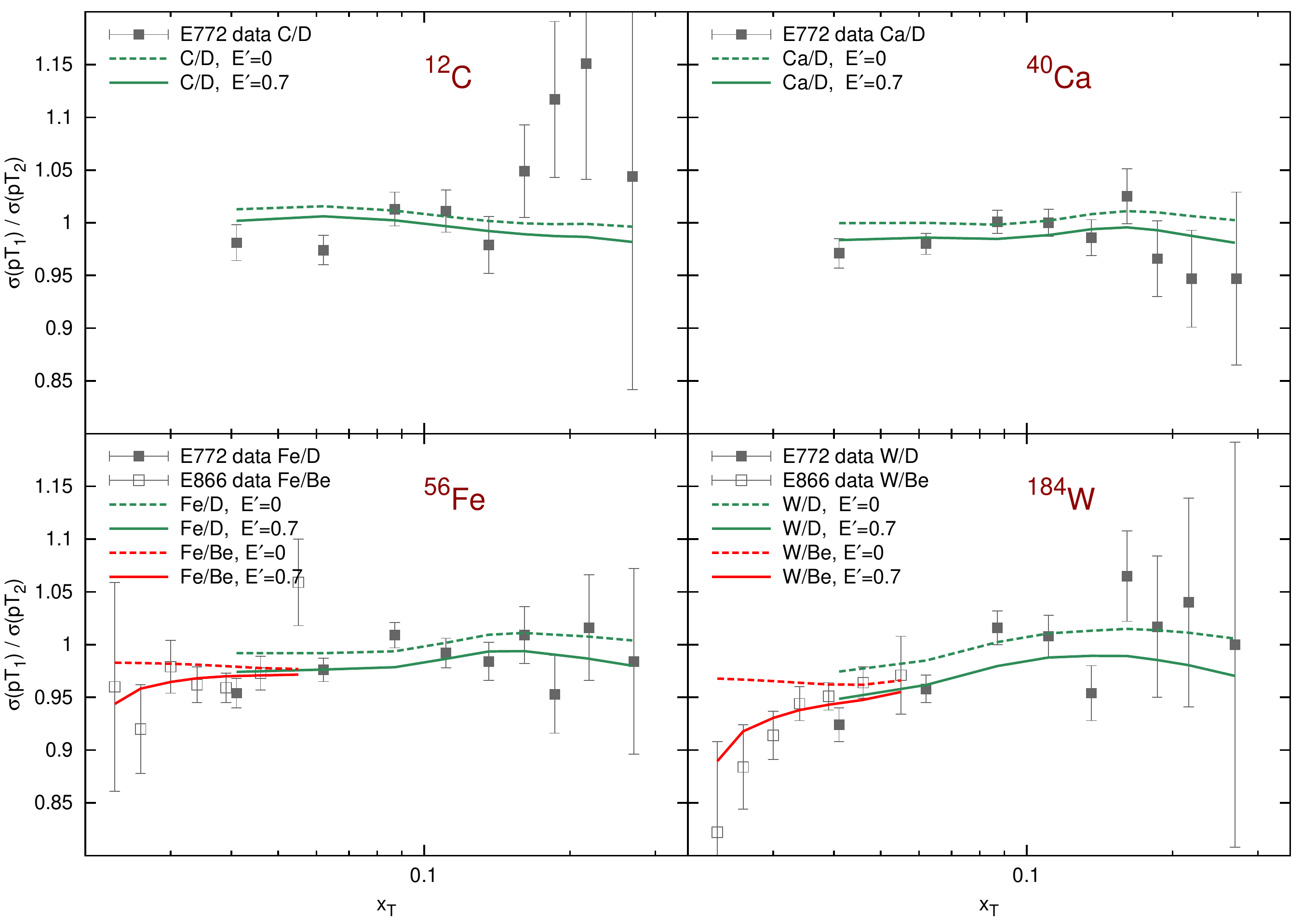}
\caption{%
Ratio of the DY reaction cross sections for different nuclei vs. $x_T$.
Data points are from the E772 experiment~\cite{E772} (\fullsquare) and E866 experiment~\cite{E866} (\opensquare).
Note that the ratio is normalized to one bound nucleon and taken relative to the deuteron for E772
and $^9$Be for E866.
The curves are the predictions of Ref.\cite{KP14} for the deuteron (green) and beryllium (red) ratios
with (\full) and without (\dashed) the projectile energy loss effect (see text and legend).
\label{fig:dy}}
\end{figure}

As discussed above, the model of Ref.\cite{KP04,KP14} predicts a significant cancellation of different nuclear effects
in the antiquark distribution in the region $x\sim 0.1$ in agreement with the nuclear DY data.
In Fig.\ref{fig:dy} we show the data along with our predictions on the ratios of the DY cross sections \cite{KP14}.
We also remark that
nuclear dependence of the DY process comes from two different sources: (i) the
modification of the nuclear target PDFs, and (ii) the initial state
interaction of the projectile particle (parton) within the nuclear environment
that causes the parton energy loss \cite{Bjorken:1982tu} before annihilation into a dimuon pair.
The rate of this effect is characterized by the parton energy loss in a nucleus per unit length $E'$
and effectively results in a change of the projectile parton Bjorken $x$ \cite{Garvey:2002sn}
which in turn modify the ratio in \eq{eq:dy_r}.

The data from the E866 experiment is shifted towards lower values of target's $x_T$ and higher
values of projectile's $x_B$ with respect to E772 data
thus falling into a region where both the shadowing and the energy loss effects become more prominent.
Our analysis \cite{KP14} indicates that the model of Ref.\cite{KP04} is in a good agreement with data
at a moderate energy loss effect.
The solid curves in Fig.\ref{fig:dy} show our predictions with $E'=0.7$~GeV/fm.
Note also that the cross section ratios are taken relative to the deuterium for E772 and beryllium for E866.
For this reason the corresponding curves in Fig.\ref{fig:dy} are not identical in the overlap region.

\section{NPDFs and total neutrino cross sections}
\label{sec:nuxsec}

For the reason of statistics the neutrino scattering data have been collected on heavy nuclei.
Thus the comparison of theory predictions with data on the neutrino cross sections
serves as an important and independent test of both the proton and the nuclear PDFs.
In the following we focus on the total cross sections as the available data
on total cross sections have typically better accuracy than the corresponding
measurements of differential cross sections and span a wide energy region.

We calculate the total cross sections by integrating the double differential inelastic
cross sections $\ud^2\sigma^{(\nu,\bar\nu)}/\ud x\ud y$ over allowed region of $x$ and $y$.
In the leading $\as$ order we have for the total cross sections off an isoscalar target
(i.e. the combination of $p+n$ or a nucleus with equal number of protons and neutrons)
\begin{equation}\label{eq:loxsec}
\frac{\sigmatot(E)}{E} = \frac{2 G_F^2 M}{3\pi(1+Q^2/M_W^2)^2}
	\left(x_0^+ \pm  x_0^-/2 + 3x_s \right),
\end{equation}
where $G_F$ is the Fermi coupling constant, $M$ and $M_W$ are the proton and the $W$-boson mass,
$x_0^\pm=\int_0^1\ud x xq_0^\pm(x,Q^2)$ is the average light-cone momentum for the isoscalar C-even ($q_0^+$)
and C-odd ($q_0^-$) quark distributions, and, similarly, $x_s$ is the light-cone momentum carried by $s$ quark
(we assume $s=\bar s$ and also neglect a small contributions from heavy quarks).
The $+/-$ sign in \eq{eq:loxsec} corresponds to $\nu/\bar \nu$ scattering.
The average invariant momentum transfer $Q$ in \eq{eq:loxsec},
although difficult to evaluate precisely,
by order of magnitude $Q^2=ME/2$.
\begin{figure}[htb]
\centering
\vspace{-3em}\includegraphics[width=0.98\textwidth]{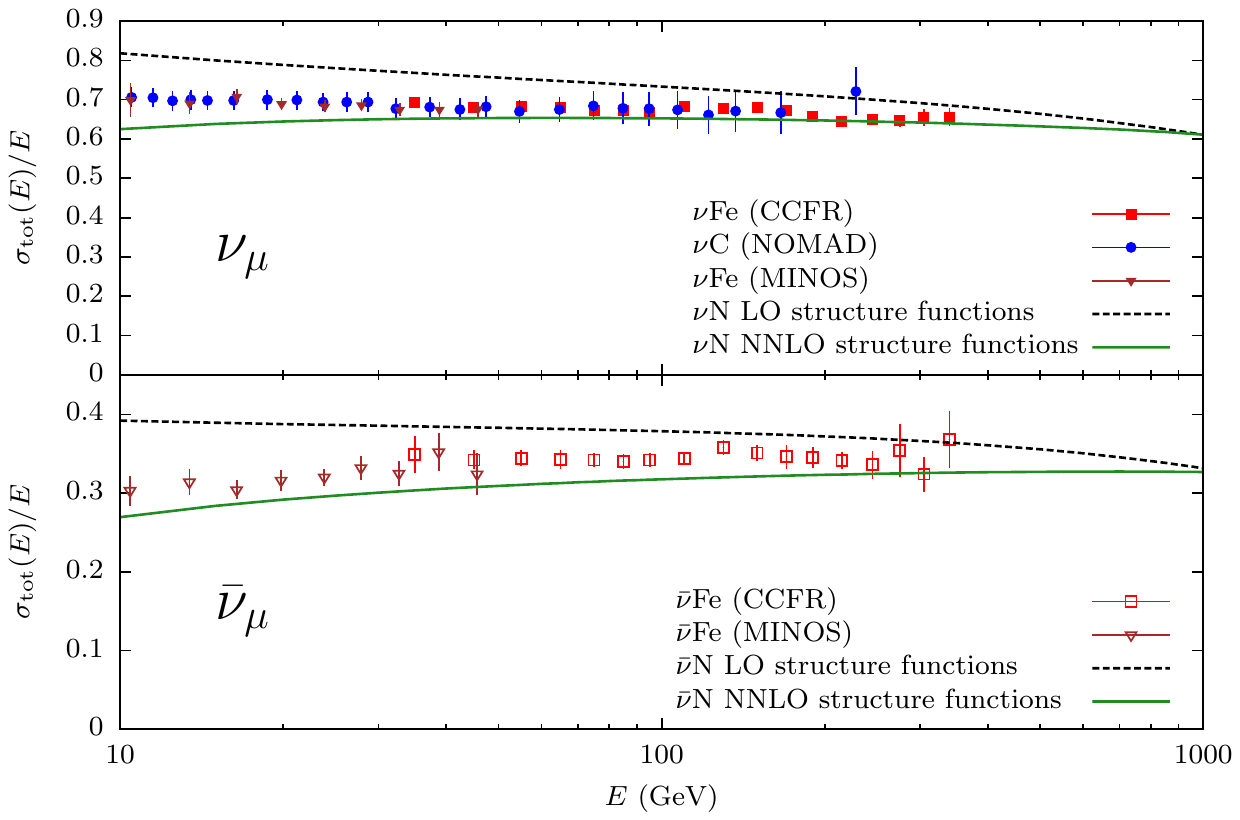}\vspace{-1em}
\caption{%
The neutrino (upper panel) and antineutrino (lower panel) total cross sections as a function of neutrino energy $E$.
The units are $10^{-38}\,\mathrm{cm}^2/\mathrm{GeV}$. The data points are the cross sections per one bound nucleon
from CCFR \cite{Seligman:1997fe}, NOMAD \cite{Wu:2007ab} and MINOS \cite{Adamson:2009ju} experiments
corrected for the neutron excess in iron.
The curves are the LO (dashed) and the NNLO (full) predictions for the isoscalar nucleon
calculated with the PDFs of Ref.\cite{Alekhin:2013nda}.
\label{fig:totxsec}
}
\end{figure}

In Fig.\ref{fig:totxsec} we show the results on the total cross section for the isoscalar nucleon 
calculated by \eq{eq:loxsec}. We also go beyond the LO approximation and numerically compute
the (anti)neutrino cross sections using the NNLO structure functions with the NNLO PDFs of Ref.\cite{Alekhin:2013nda}.

A few comments are in order.
The ratio $\sigmatot(E)/E$ is almost a constant in the considered energy region.
A weak energy dependence comes through $Q^2$ dependence of $W$-boson propagator and
the parton distributions (structure functions).
A higher order $\as$ corrections on the structure functions result in a overall negative
correction to the total cross section reducing significantly the LO cross section
in the energy region up to a few hundred GeV.
It is worth noting that this effect leads to a better agreement with data.

\begin{figure}[htb]
\includegraphics[width=0.5\linewidth]{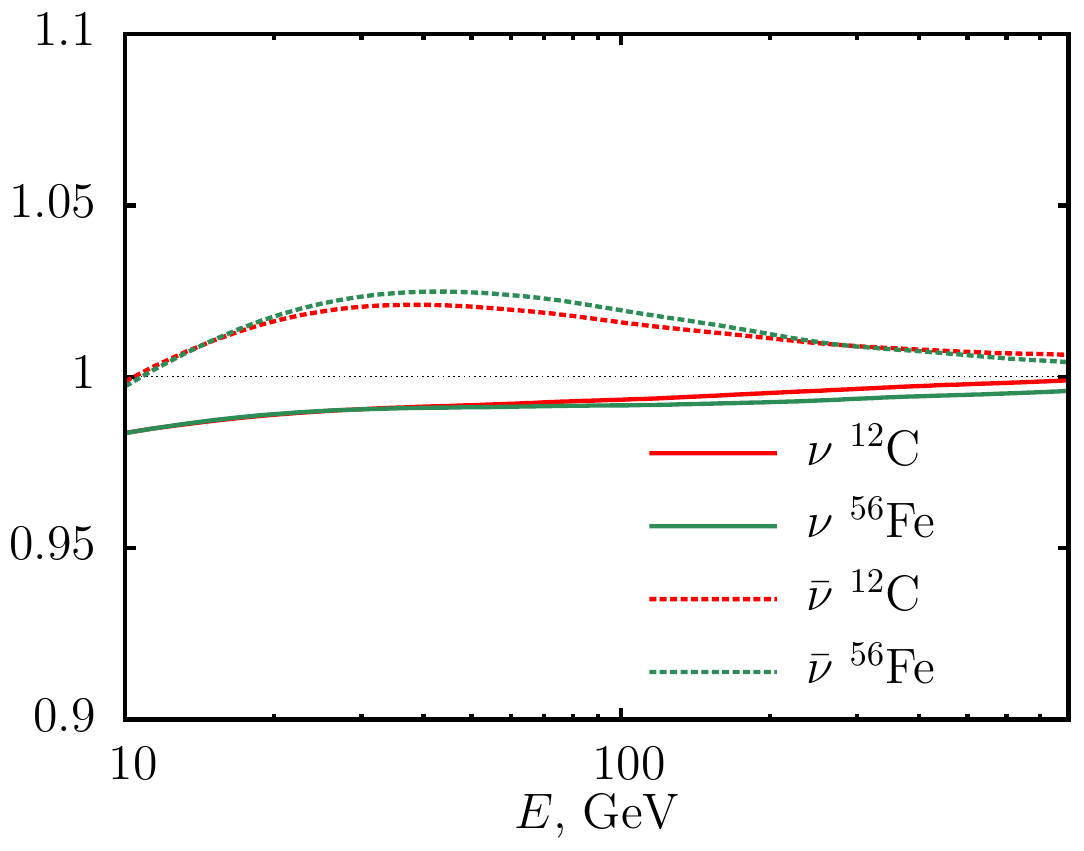}%
\includegraphics[width=0.5\linewidth]{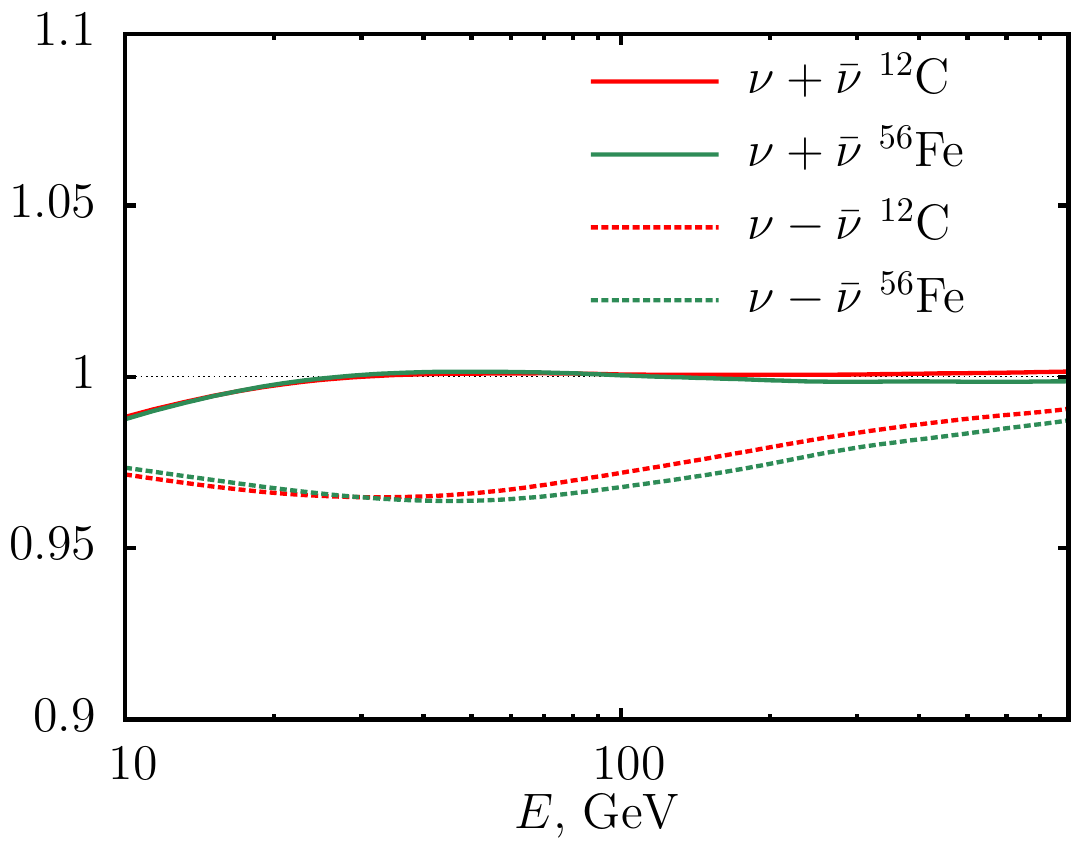}\vspace{-1em}
\caption{%
Left panel:
The ratios $R^{\nu}_A$ (solid curve) and $R^{\bar\nu}_A$ (dashed curve)
calculated for $^{12}$C (red) and $^{56}$Fe (green) nuclei as a function of neutrino energy $E$.
Right panel:
Similar ratios for the C-even $\nu+\bar\nu$ (solid) and C-odd $\nu-\bar\nu$ (dashed)
combinations of the total cross sections.}
\label{fig:xsec_rat}
\end{figure}

In Fig.\ref{fig:totxsec} we show the data from CCFR \cite{Seligman:1997fe},
NOMAD \cite{Wu:2007ab} and MINOS \cite{Adamson:2009ju} neutrino experiments.
Note, however, those data
were taken on different nuclear targets, $^{56}$Fe for CCFR and MINOS and $^{12}$C for NOMAD.
Therefore, in order to make a meaningful comparison,
it is important to address the nuclear dependence of the (anti)neutrino cross sections.
%
For this reason we discuss the normalized ratios of the nuclear and the proton and neutron
neutrino cross sections
$R^\nu_A=\sigma^\nu_A/(Z\sigma^\nu_p +N\sigma^\nu_n)$
and a similar ratio $R^{\bar\nu}_A$ for antineutrino.
In Fig.\ref{fig:xsec_rat} (left panel) we show the ratios $R^\nu_A$ and $R^{\bar\nu}_A$
calculated for $^{12}$C and $^{56}$Fe nuclei using the NNLO structure functions of Ref.\cite{Alekhin:2013nda}
within the approach outlined above (for more detail see Refs.\cite{KP07,KP14}).
We observe that in considered energy region
the rate of nuclear effects is about 2\% at maximum for both the neutrino
and the antineutrino and gradually decreasing with energy.
Note also that the ratios $R^\nu_A$ and $R^{\bar\nu}_A$ show almost no nuclear dependence that justifies
the comparison of total neutrino cross sections for different nuclear targets in Fig.\ref{fig:totxsec}.
%
%

Figure~\ref{fig:xsec_rat} indicates a suppression of the  neutrino-nuclear total cross section
while the antineutrino cross section is somewhat enhanced in nuclei
(the latter in turn further improves agreement with $\bar\nu$ data in Fig.\ref{fig:totxsec}).
These two effects cancel each other for the C-even combination $\nu+\bar\nu$
of the cross sections as illustrated in the right panel of Fig.\ref{fig:xsec_rat}.
On the other hand, the nuclear correction to the C-odd combination $\nu-\bar\nu$ is somewhat enhanced.


It is instructive to compare the nuclear corrections for the total cross sections
with those for PDFs (see Fig.\ref{fig:npdfs} and \ref{fig:emc-effect}).
We note in this context a direct relation between the total neutrino cross sections and average
light-cone momenta of different quark PDFs in \eq{eq:loxsec}.
The nuclear modification of average light-cone momenta have much smaller magnitude compared to a typical
amplitude of nuclear effects for PDFs shown in Fig.\ref{fig:npdfs} and \ref{fig:emc-effect}.
This is because of a cancellation between nuclear corrections of different sign in different
kinematic regions in the integral over the Bjorken $x$.
The isoscalar $\nu+\bar\nu$ combination of the total cross sections is proportional to
the total quark and antiquark light-cone momentum in the target.
Thus a cancellation of nuclear effects in $\sigma(\nu+\bar\nu)$ suggests that the overall light-cone momentum
of quarks and antiquarks does not change in nuclei.
Referring to the light-cone sum rule we also conclude that
the average gluon light-cone momentum is not modified in nuclei.
However, discussion of associated problems goes beyond of the scope of the present article and will be given elsewhere.

\section{Summary}
\label{sec:sum}

In summary, we reviewed a semi-microscopic model of nuclear PDFs
addressing a few different mechanisms of modification of PDFs in nuclear environment.
A number of applications were discussed including charged-lepton DIS,
proton-nuclear DY reaction
and calculation of total (anti)neutrino-nuclear cross sections in charged-current interaction.


I thank R. Petti for fruitful collaboration on the topics of this paper
and A. Kataev for motivating discussions.
I am grateful to the Organizing Committee of ACAT 2016 Workshop for support and
for creating a friendly and working environment during the meeting.

The work was supported by the Russian Science Foundation grant No.~14-22-00161.

\end{document}